\documentstyle[prd,aps,eqsecnum,epsf]{revtex}

\def\no{\nonumber \\}
\def\le{\biggl \{}
\def\ri{\biggr \}}    

\begin{document}

\title{Pad\'e approximants for
truncated post-Newtonian neutron star models  }

\author{Anshu Gupta$^{1}$, A. Gopakumar$^{2,3}$, Bala  R. Iyer$^{1}$ 
and Sai Iyer$^2$}

\address{ $^1$Raman Research Institute, C.V. Raman Avenue,
Sadashivanagar, Bangalore 560080,
India\\
$^2$Physical Research Laboratory, Navrangpura, Ahmedabad 380 009,
India\\
$^3$Department of Physics and McDonnell Center for the Space Sciences,
\\ Washington University, St. Louis, MO 63130, U.S.A}

\date{\today}

\maketitle

\begin{abstract}
 Pad\'e approximants to truncated post-Newtonian 
 neutron star models are constructed.
The Pad\'e models converge faster to the general relativistic (GR)
 solution  than the  truncated  post-Newtonian  ones. 
The evolution of initial data  using the Pad\'e models 
  approximates better the evolution of full GR
 initial data  than the truncated Taylor  models. 
In the absence of full GR initial data (e.g., for neutron
star binaries or black hole binary systems),
Pad\'e initial data could be a better option than the
straightforward truncated  post-Newtonian (Taylor) initial data. 
\end{abstract}


\section{Introduction}

Compact binary systems of neutron stars and black holes inspiralling
under gravitational radiation reaction are one of the most promising
sources of gravitational waves for kilometre arm length
interferometric gravitational wave detectors like the LIGO and VIRGO.
The inspiral phase is best described by a post-Newtonian approximation
which should eventually break down and the final merger and coalescence
may be only accessible via numerical integration of Einstein's equations.
A  major obstacle in the numerical studies of such
systems is the  non-availability of physically satisfactory initial data. 
Given the constraints on computational resources, one would like to start
the numerical integration as close as possible to the final coalescence
phase, using the initial data obtained by matching on to the known
analytical results of inspiral.
One of the suggestions in this direction is to  use the analytical
post-Newtonian results of inspiral to provide initial data 
(e.g. \cite{mg8}-\cite{bd00}) for the
numerical  integration  of  the fully  general relativistic system. 

Before attempting to apply the above strategy
to the  complicated   compact binary  system, 
as a preliminary step, Shinkai \cite{shinkai} has constructed a single 
neutron star model using the post-Newtonian approach and concluded
 that the truncated  second order
 post-Newtonian  approximation is close enough to describe
a general relativistic single star.
 The truncated post-Newtonian series used above
 is essentially a Taylor expansion in the three small parameters
 $A=p/(\rho_t c^2)$, $B= 4\pi p r^3/(mc^2)$
and $C=2Gm/(rc^2)$  where 
$p$ is the pressure, $\rho_t$, the total energy density
of the system, $r$ the radial coordinate, and
 $m(r)$  the mass contained in a sphere of radial coordinate $r$.
We shall refer to the post-Newtonian truncated models alternatively
as {\it Taylor models}. Recently,  
in a related context of gravitational wave phasing, 
the slow convergence of  straightforward
Taylor approximants has been critically investigated\cite{dis1}\cite{dis2}. 
It was shown that new approximants, with much improved convergence properties,
may be constructed
for gravitational wave data analysis applications using,
as an important tool, Pad\'e techniques to estimate the relevant
functions from only the first few terms in their perturbative
post-Newtonian expansion.
This approach has been systematically extended in a series
of publications \cite{bd99}\cite{djs99} 
and most recently has been used to  go beyond the adiabatic approximation
to inspiral  and provide an analysis of the transition from the inspiral to
the plunge in binary black hole coalescences \cite{bd00}. This work also
provides for the first time initial dynamical data (positions and
momenta) for binary black holes starting to plunge, so that
there is less than an orbit left to evolve.
In view of this experience, in this report, we   construct Pad\'e 
approximants   to   the   truncated  post-Newtonian (Taylor)
neutron star models discussed by Shinkai \cite{shinkai} and investigate 
their performance.
 We show that the Pad\'e models are  better than the truncated post-Newtonian
models and converge faster to   the exact  general relativistic solution. 
Further, we also  show that the evolution of a  general relativistic (single) 
star is described much better by Pad\'e initial data as compared to the 
truncated Taylor initial data of the same order. 

In the next section we discuss the TOV equation and the truncated 
post-Newtonian models.
In  Section III we  introduce the standard   Pad\'e approximant,
discuss its applicability to the present problem,
 and  adapt it  to construct   an appropriate   generalisation 
 for the case of  3-small parameters. 
Pad\'e models corresponding to the `Taylor' truncated models
are constructed.
In the last  section  we discuss the results and summarise our  conclusions.
The Appendix lists the more involved formulas for the 3PN Pad\'e
models.

\section{TOV equation and  truncated post-Newtonian (Taylor) neutron star models}

In general relativity the metric of a spherically symmetric static star
can be written as
\begin{equation}
ds^2 = -e^{2 \nu (r)} dt^2 + e^{2 \lambda (r) } dr^2 + r^2 (d\theta^2 + r^2 \sin^2\theta
d\phi^2),
\label{eq:met}
\end{equation}
and the equations of hydrostatic equilibrium - the Tolman-Oppenheimer-Volkoff
(TOV)  equations -  
 obtained (in geometrised units: $G = c = M_{\odot} = 1$) from 
the Einstein field equations, for a given fluid distribution
 specified by an adiabatic equation of state $p = p (\rho_t)$, 
are given by:
\begin{eqnarray}
\label{eq:dmr}
\frac{d m}{d r} &=& 4 \pi r^2 \rho_t,\\
\label{eq:dpr}
\frac{d p}{d r} &=& - \frac{m \rho_t}{r^2} \left(1 + \frac{p}{\rho_t}\right) 
\left(1 + \frac{4 \pi p r^3}{m}\right) \left(1 - \frac{2 m}{r}\right)^{-1},\\
\label{eq:dnr}
\frac{d \nu}{d r} &=& \frac{m}{r^2} \left(1 + \frac{4 \pi p r^3}{m}\right) 
\left(1 - \frac{2 m}{r}\right)^{-1}.
\end{eqnarray}
$m$ is given  in terms of metric component $\lambda$ as 
\begin{equation}
e^{2 \lambda} = \left(1-\frac{2 m}{r}\right)^{-1}. 
\end{equation}
The above set of equations are 
integrated from the center ($r=0$) to the boundary of the star, with the initial 
conditions $m(r=0)=0; \rho_t(r=0) = \rho_{t_c}$ and $\nu{(r=0)}=\nu_c$.
$\nu_c$ is rescaled appropriately such that it matches with the exterior
Schwarzschild solution at the boundary.  The radius of the
star, $R$, is  characterised as the radius
 where density $\rho_t(r=R)$ drops to 
$10^6$ gm/cc (approximately ${\cal O} (10^{-10})$ in geometrised units).

\begin{figure}
\epsfxsize=6in \epsfbox{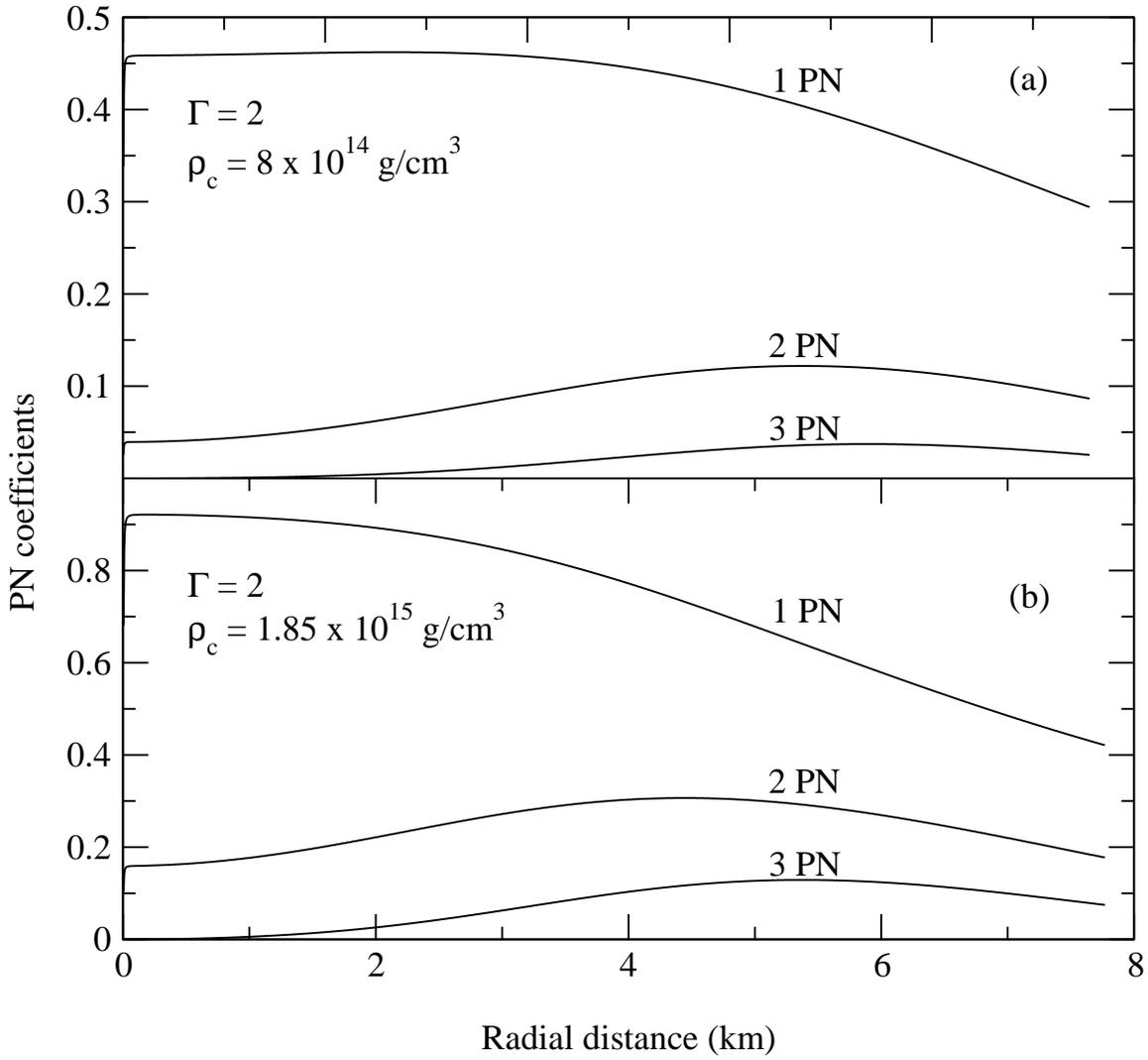}
\bigskip

\caption{Post-Newtonian coefficients appearing in  Eq. (\ref{eq:dpr3}) as a 
function of the radial distance (in units of km). The models are for a 
polytropic equation of state with intermediate stiffness ($\Gamma = 2$) 
with the central density $\rho_c = 
8$ x $10^{14}$ g/cm$^3$ (panel (a)) and $1.85$ x $10^{15}$ g/cm$^3$ (panel (b)).
The latter is a model which is very close to the maximum mass limit.}
\label{fig:r1}
\end{figure}

Schematically, equations (\ref{eq:dpr}) and (\ref{eq:dnr}) can be written as
\begin{eqnarray}
\label{eq:dpr2}
\frac{dp}{dr} &=& -\frac{m \rho_t}{r^2} (1+A) (1+B) (1 -C)^{-1}\,,\\
\label{eq:dnr2}
\frac{d \nu}{d r} &=& \frac{m}{r^2}(1+B) (1-C)^{-1}.
\end{eqnarray}
 Assuming that $A$, $B$ and $C$ are of comparable orders of smallness
and making a  Taylor series expansion in $A$, $B$, and $C$ around the origin,
the above equations to  third   post-Newtonian  order then  yield, 
\begin{eqnarray}
\label{eq:dpr3}
\frac{dp}{dr} &=& -\frac{m \rho_t}{r^2} (1 + \underbrace{ A +B +C}_{1 PN}
+ \underbrace{AB +BC +CA + C^2}_{2 PN} \nonumber \\
& &+ \underbrace{ABC + AC^2 + BC^2 +C^3}_{3 PN} + {\cal O}(4)\cdots)\,,\\
\label{eq:dnr3}
\frac{d \nu}{d r} &=& \frac{m}{r^2}(1 + \underbrace{B +C}_{1 PN}
+ \underbrace{BC + C^2}_{2 PN}+ \underbrace{BC^2 +C^3}_{3 PN} + 
{\cal O}(4)\cdots)\,,
\end{eqnarray}
 where the post-Newtonian order of the relevant terms is indicated by
  the PN label under the braces. 
These equations describe the truncated post-Newtonian model  
\cite{shinkai}.
Before proceeding ahead, we must verify that the above truncation is 
consistent and meaningful in the present problem of neutron stars.
For a range of models, we evaluate the successive combinations 
that appear in 
Eqs. (\ref{eq:dpr3}) and (\ref{eq:dnr3}), using the exact TOV equations. 
Figure~\ref{fig:r1} shows the combinations appearing in Eq. (\ref{eq:dpr3}) 
for two such models with an equation of state of intermediate stiffness, 
one which is very close to the maximum mass limit and another, a little
further down the stable branch.
From the Figure it is clear that   numerically the successive combinations
are perturbatively smaller to be consistently  denoted as 
1PN, 2PN and 3PN contributions.

\section{Pad\'e approximants to the truncated post-Newtonian (Taylor) models}

 Given a Taylor series of order $n$ in one expansion parameter $x$
\begin{equation}
S_n(x) = 1 + a_1 x + \cdots +a_n x^n,
\label{eq:tayl}
\end{equation}
its  Pad\'e approximant is the ratio of rational functions,
\begin{equation}
P^m_k (x) = \frac{N_m (x)}{D_k (x)},
\end{equation}
 satisfying the condition
\begin{equation}
T_n[P^m_k (x)] \equiv S_n (x);\;\;\;\; k+m = n.
\label{eq:pade}
\end{equation}
In the above,
$N_m$ and $D_k$ are polynomials in $x$ of order $m$ and $k$ respectively and 
$T_n$ denotes the operation of expanding a function in Taylor series upto
order $n$.

 A convenient choice   of Pad\'e approximants is its diagonal (or nearly 
diagonal) continued fraction form \cite{BO} i.e., $P^m_m$ when $n=2m$ (even)
and $P^m_{m+1}$ when $n=2m+1$ (odd).  
For instance,
\begin{equation}
P^1_2 = \frac{c_0}{1+\displaystyle{\frac{c_1 x}
{1+\displaystyle{\frac{c_2 x}{1+c_3 x}}}}}\,,
\label{eq:2pd1}
\end{equation}
where the $c_i$'s are determined in terms of the Taylor coefficients
$a_i$ by 

\begin{equation}
c_0 = 1,\;\; c_1 = -a_1,\;\; c_2 = \frac{a_1^2 - a_2}{a_1},\;\; 
c_3=\frac{a_1 a_3-a_2^2}{a_1(a_1^2-a_2)}\,.
\label{eq:one}
\end{equation}

\begin{figure}
\epsfxsize=6in \epsfbox{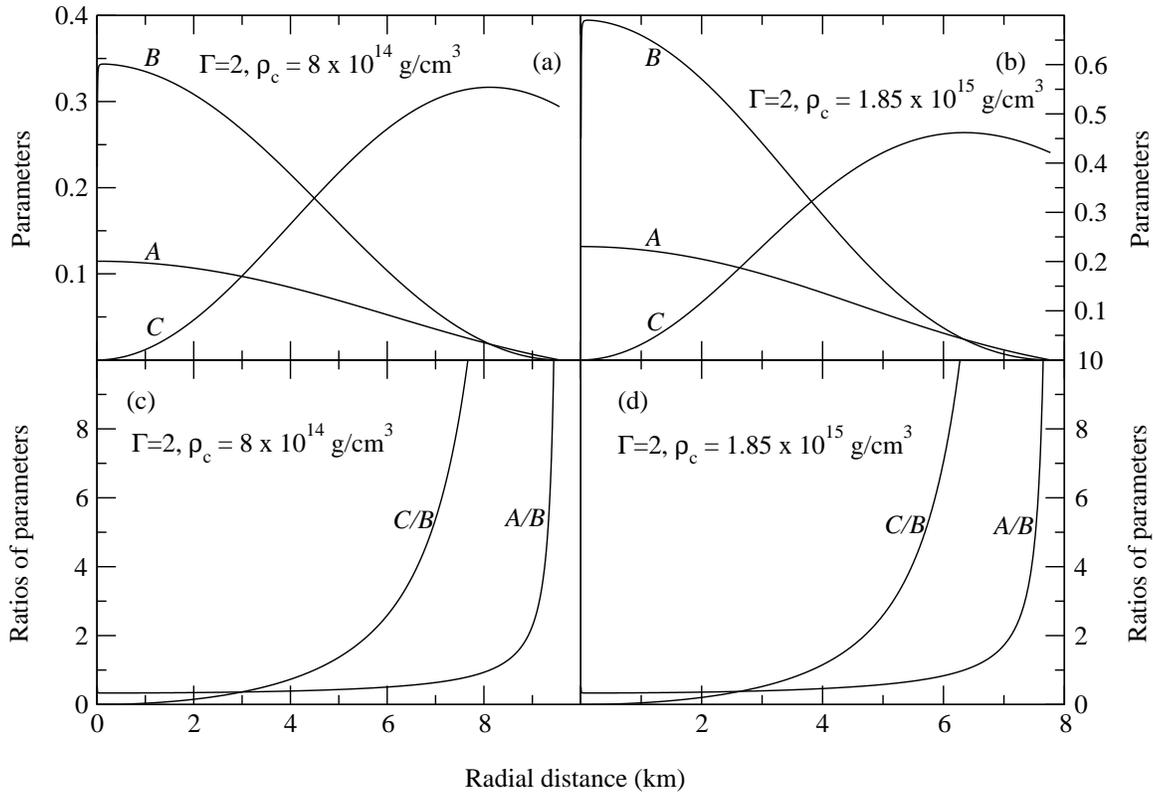}
\bigskip

\caption{Parameters $A, B, C$ and the ratios of parameters $A/B, C/B$ 
as a function of radial distance. From the panels (c) and (d), we see the 
weak dependence of the ratio $C/B$ on $r$ (the radial distance), though
$A/B$ is almost constant all through, except near the surface.}
\label{fig:r2}
\end{figure}

To construct the Pad\'e approximant of the truncated post-Newtonian
models discussed in the previous section, we must proceed carefully.
In the problem under discussion, in general, there are three independent small
parameters $A$, $B$, $C$, of the same order of smallness.
Given  this  system  of approximate differential equations containing
three independent small parameters, 
 can  we construct an associated Pad\'e like approximation with faster
 and improved convergence?
Assume for a moment, that since $A$, $B$, $C$ are of the same order of
smallness, we decide to use as the variable of expansion the parameter
$B$ that takes the maximum value in the entire range of integration.
The first order Taylor term $A+B+C$ would be rewritten as
$ B (1+A/B + C/B)$.
If we treat only $B$ as the independent variable	then its
associated coefficient would be	 $1+A/B + C/B$.
If $A/B$ and $C/B$ have only a weak dependence on $r$,
we can indeed use a simpler Pad\'e form with only one variable and
this should suffice.  
We call this the one parameter Pad\'e form for the TOV equations.
In this case, the relevant equations  at  2PN are the following:\footnote{
Identical equations obtain if one uses any of $A$, $B$ or $C$ as the
expansion parameter. It is also
  equivalent to introducing by hand a small parameter
say $\varepsilon$ in terms of which we (Taylor or Pad\'e) expand,
treating as coefficients the associated  $A, B, C$ dependence.} 
\begin{mathletters}
\begin{eqnarray}
\frac{dm}{dr} &=& 4 \pi r^2 \rho_t\,,\\
\label{eq:dpr_1p}
\frac{dp}{dr} &=& -\frac{m \rho_t}{r^2} \frac{1}{{\cal F}_2}\,,\\
\label{eq:dnur_1p}
\frac{d \nu}{d r} &=& \frac{m}{r^2} \frac{1}{{\cal G}_2}\,,\\
{\rm where},\;\; {\cal F}_2&=&\frac{A+B+C-AB-BC-CA-C^2}
{A+B+C+A^2+AB+CA+B^2+BC}\,,\\
{\rm and},\;\; {\cal G}_2&=&
\frac{B+C-BC-C^2} {B+C+B^2+BC}\,.
\end{eqnarray}
\end{mathletters}
(The more involved equations\footnote{The explicit 
forms of ${\cal F}_2, {\cal G}_2$ and the corresponding
3PN forms in the Appendix are exibited for analytical completeness. 
In our numerical computation however, we directly substitute 
Eqs. (\ref{eq:2pd1}) and (\ref{eq:one}) in Eqs. (\ref{eq:dpr_1p}) and 
(\ref{eq:dnur_1p}).}
at 3PN are listed in the Appendix).
To examine more quantitatively, the validity of the above treatment,
 we consider, as before, a few models and compute the values of 
$A$, $B$ and $C$  and the associated ratios of $A/B$ and $C/B$.
The results are displayed in Figure~\ref{fig:r2}. From the figure
it is clear, that as required,  $A$, $B$, $C$ are approximately
of the same order of smallness.
However, if one looks at the ratio of  $C/B$, one finds that
there is a regime where the $r$-dependence is not very  weak.
Consequently, we are wary of using only  the usual straightforward
 one parameter Pad\'e form discussed above and proceed as follows.
 We generalise the usual construction for the one parameter Pad\'e
 to the situation where there exist, not one, but three small parameters.
The advantage of using	a three  parameter form is that its
`coefficients' are pure numbers with no explicit  `$r$-dependence'.
By treating $A$, $B$, $C$ as independent variables, we avoid the explicit
 issue of the `weak $r$-dependence' of the associated coefficients.\footnote{
Indeed, one cannot relax this requirement
 in our three parameter construction either; however, this is only
   implicit in the fact	 that $A$, $B$, $C$ are of the same order of smallness.}

To this end, we start with   the most general second order 
post-Newtonian accurate polynomial in  three small parameters $A$, $B$
and $C$ as
\begin{eqnarray}
T_2 & = & 1 + t_A A + t_B B + t_C C + t_{A A} A^2 + t_{B B} B^2 + t_{C C} C^2 \nonumber \\
    & & + t_{A B} A B + t_{B C} B C + t_{C A} C A \,.
\label{eq:tayl2}
\end{eqnarray}
The associated  Pad\'e approximant in continued fraction 
form may be chosen as
\begin{equation}
\label{eq:2pd3a}
P^1_1 = \frac{1}{F_2}\,,
\end{equation}
where
\begin{eqnarray}
\label{eq:2pd3}
F_2 & = & 1 +\frac{p_1 A}{1+p_{1 1} A + p_{1 2} B + p_{1 3} C} + \nonumber \\
& &\frac{p_2 B}{1+p_{2 1} A + p_{2 2} B + p_{2 3} C} 
+ \frac{p_3 C}{1+p_{3 1} A + p_{3 2} B + p_{3 3} C}\,. 
\end{eqnarray}
By matching coefficients of the Taylor expansion of $P^1_1$,
 Eq. (\ref{eq:2pd3a}), and $T_2$,  Eq. (\ref{eq:tayl2}),  the three coefficients
$p_1$, $p_2$  and $p_3$ are uniquely determined as, 
\begin{equation}
p_1 = -t_A; \;\;\; p_2 = -t_B;\;\;\; p_3 = -t_C.
\end{equation}
The remaining nine  coefficients $p_{i j}$, where $i, j =1, 2, 3$,
are not uniquely determined; they are solutions to   
 the following  (under-determined) system of  six equations 
\begin{mathletters}
\begin{eqnarray}
\label{eq:p11}
t_A (t_A - p_{1 1}) &=& t_{A A}\,,\\
\label{eq:p22}
t_B (t_B - p_{2 2}) &=& t_{B B}\,,\\
\label{eq:p33}
t_C (t_C - p_{3 3}) &=& t_{C C}\,,\\
\label{eq:p12}
t_A p_{1 2} + t_B p_{2 1} &=& 2 t_A t_B - t_{A B}\, ,\\
\label{eq:p23}
t_B p_{2 3} + t_C p_{3 2} &=& 2 t_B t_C - t_{B C}\, ,\\
\label{eq:p31}
t_C p_{3 1} + t_A p_{1 3} &=& 2 t_C t_A - t_{C A}\,.
\end{eqnarray}
\end{mathletters}
If $t_A\neq 0$,  $t_B\neq 0$,  $t_C \neq 0$,  
$p_{11}$, $p_{22}$ and  $p_{33}$,  
can also be uniquely determined
using Eqs. (\ref{eq:p11}), (\ref{eq:p22}) and (\ref{eq:p33})  respectively.
We have
\begin{mathletters}
\begin{eqnarray}
\label{eq:pii1}
p_{11}&=&\frac{t_A^2-t_{AA}}{t_A},\\
\label{eq:pii2}
p_{22}&=&\frac{t_B^2-t_{BB}}{t_B},\\
\label{eq:pii3}
p_{33}&=&\frac{t_C^2-t_{CC}}{t_C}.
\end{eqnarray}
\end{mathletters}
The remaining six  non-diagonal terms cannot be  uniquely determined
from the remaining three equations Eq. (\ref{eq:p12}) - (\ref{eq:p31}) 
without further input.
The natural requirement that all Pad\'e coefficients $p_{ij}$
 contributing to a particular Taylor term,
contribute equally, leads one to the following symmetry  choice  
\begin{equation}
 p_{12}=p_{21},\;\; 
 p_{23}=p_{32},\;\; 
 p_{13}=p_{31}. 
\end{equation}
One can then   uniquely determine all the required coefficients 
and finally obtain
\begin{mathletters}
\begin{eqnarray}
\label{eq:pij1}
p_{12}=p_{21}&=&\frac{2t_At_B-t_{AB}}{t_A+t_B},\;\;{\rm if }\;\;\; t_A+t_B\neq 0,\\
\label{eq:pij2}
p_{23}=p_{32}&=&\frac{2t_Bt_C-t_{BC}}{t_B+t_C},\;\;{\rm if }\;\;\; t_B+t_C\neq 0,\\
\label{eq:pij3}
p_{31}=p_{13}&=&\frac{2t_Ct_A-t_{CA}}{t_C+t_A},\;\;{\rm if }\;\;\; t_C+t_A\neq 0.
\end{eqnarray}
\end{mathletters}
Since the form of Eq. (\ref{eq:dpr3}) is equivalent to Eq. (\ref{eq:tayl2}) and
 Eq. (\ref{eq:dnr2}) has a  less general form containing only 2-parameters $B$ and $C$, 
the second order Pad\'e approximant to the second order
truncated TOV equations may be written as
\begin{eqnarray}
\frac{dm}{dr} &=& 4 \pi r^2 \rho_t\,,\\
\frac{dp}{dr} &=& -\frac{m \rho_t}{r^2} \frac{1}{F_2}\,,\\
\frac{d \nu}{d r} &=& \frac{m}{r^2} \frac{1}{G_2}\,,
\end{eqnarray}
where $F_2$ is given by  Eq. (\ref{eq:2pd3}) and $G_2$ is obtained
by substituting $A=0$ in $F_2$ .  A comparison of 
Eqs. (\ref{eq:dpr3}) and (\ref{eq:tayl2})
yields 
\begin{mathletters}
\begin{eqnarray}
t_A=t_B=t_C=t_{AB}=t_{BC}=t_{CA}=t_{CC}=1,\\
t_{BB}=t_{AA}=0,
\end{eqnarray}
\end{mathletters}
 so that the $p_{ij}$ (Eqs. (\ref{eq:pii1})-(\ref{eq:pii3})
 and (\ref{eq:pij1})-(\ref{eq:pij3}))
 in this case  become
\begin{mathletters}
\begin{eqnarray}
p_{11} = p_{22} = 1,\;\;p_{33}=0\,,\\
p_{12}= p_{23}= p_{13} = \frac{1}{2}\,.
\end{eqnarray}
\end{mathletters}
The associated forms of $F_2$ and $G_2$ are finally  given by
\begin{eqnarray}
\label{eq:tovf1}
F_2 &=& 1 -\frac{A}{1+A + \frac{B + C}{2}} -\frac{B}{1+B + \frac{A + C}{2}}
- \frac{C}{1+\frac{A+B}{2}}\;, \\
\label{eq:tovf2}
G_2 &=& 1 -\frac{B}{1+B + \frac{C}{2}} - \frac{C}{1+\frac{B}{2}}\; .
\end{eqnarray}
Equations (\ref{eq:tovf1}) and (\ref{eq:tovf2}) define the three parameter
2PN Pad\'e model associated with the 2PN truncated model discussed in 
\cite{shinkai}. The 3PN three parameter Pad\'e model may be similarly 
obtained but since the expressions are lengthier, we list them in the Appendix
together with the one parameter 3PN Pad\'e model.

\section{Results and Discussion}

\begin{figure}
\epsfxsize=6in \epsfbox{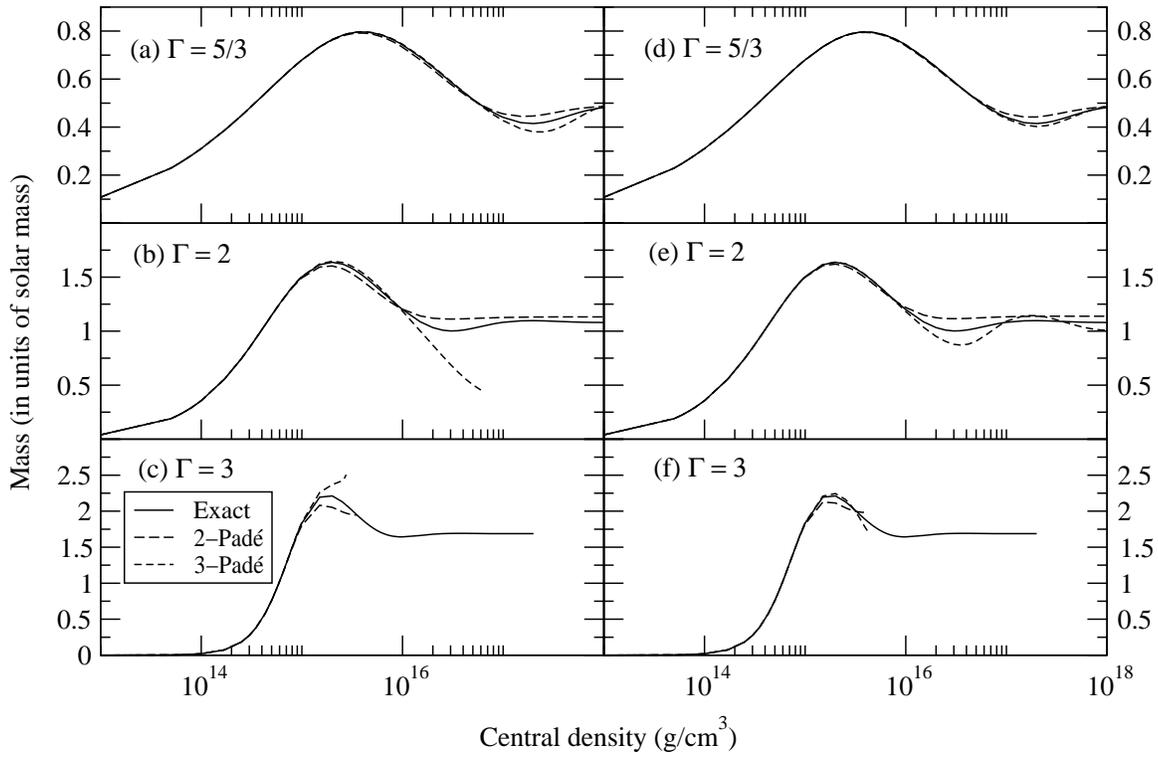}
\bigskip

\caption{A comparision of the two different Pad\'e aprroximants 
constructed here: the one parameter Pad\'e in
panels [d-f] and three parameter Pad\'e in
panels [a-c]. It is clear that in the stable region
both models give very similar results.}
\label{fig:r3}
\end{figure}

We consider the polytropic equation of state $p = K \rho^{\Gamma}$ choosing the
same models as studied by Shinkai \cite{shinkai} 
i.e., $\Gamma = 5/3, 2$ and $3$ with $K$ (the polytropic constant) $= 4.35,
10^2$ and $10^5$, respectively. 
Before presenting the final results, we have compared the performance of
the one parameter Pad\'e models with three parameter Pad\'e models
and display in Figure~\ref{fig:r3} a typical comparison.
From the Figure it is clear that both models yield similar
numerical performance and the difference, if any, appears in the
region where anyway these models are not recommended.
Consequently, all our Figures refer to the one parameter Pad\'e
models though we have verified that the three parameter Pad\'e
 models also yield similar results.

\begin{figure}
\epsfxsize=6in \epsfbox{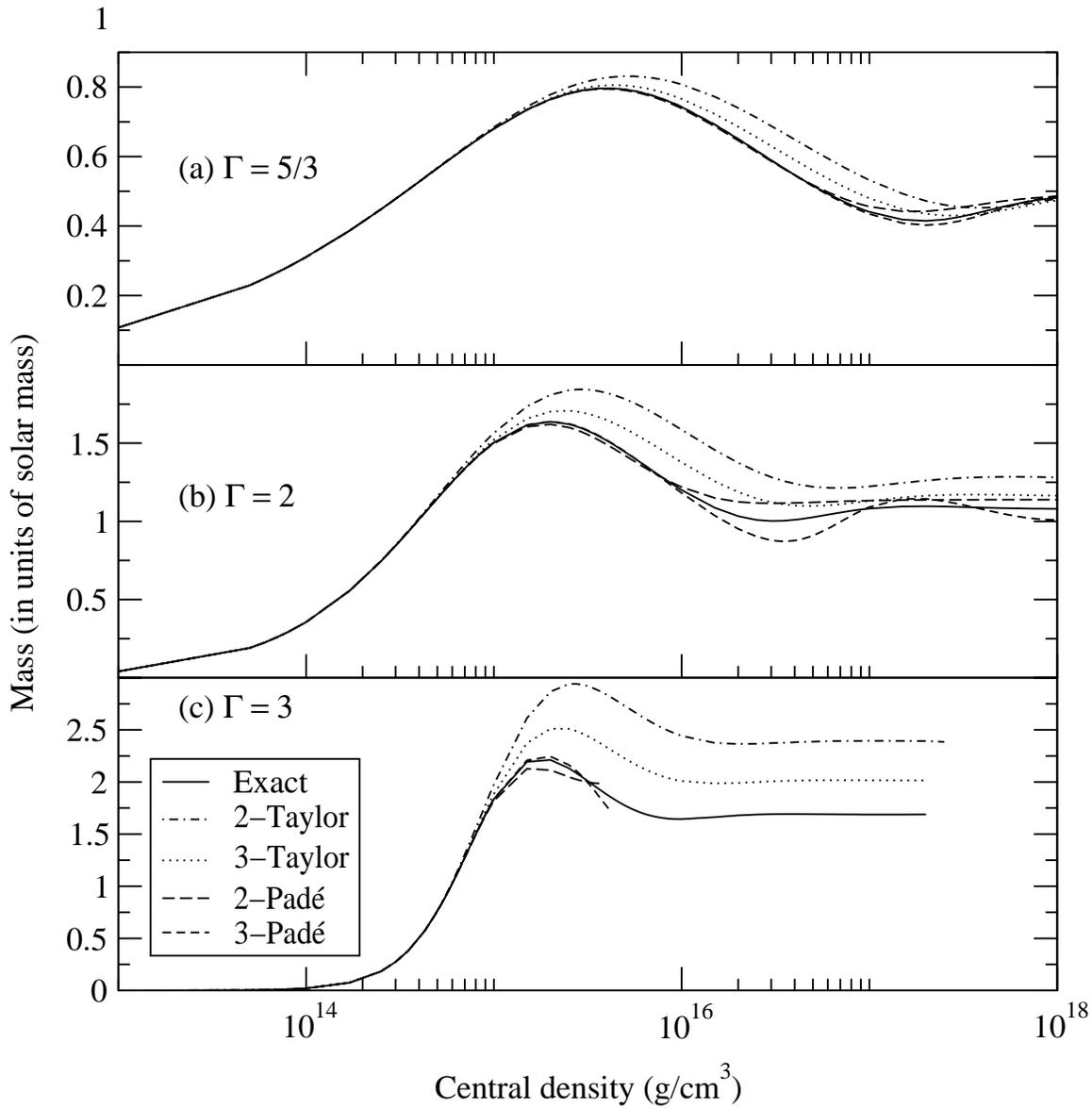}
\bigskip

\caption{Total mass (in units of $M_{\odot}$) as a function of
  central density ($\rho_c$) (in units of g/cm$^3$) for three values of
  $\Gamma$. Displayed here and in all subsequent figures are the one 
  parameter Pad\'e models. The three parameter Pad\'e models also give 
  very similiar results. Pad\'e approximants do extremely well as compared 
  to Taylor truncated models upto maximum mass limit i.e., for all the
  stable TOV models (models to the left of the first extrema). For the
  behaviour in the region beyond this extremum, see the discussion in
  section 4.}
\label{fig:f1}
\end{figure}

\begin{figure}
\epsfxsize=6in \epsfbox{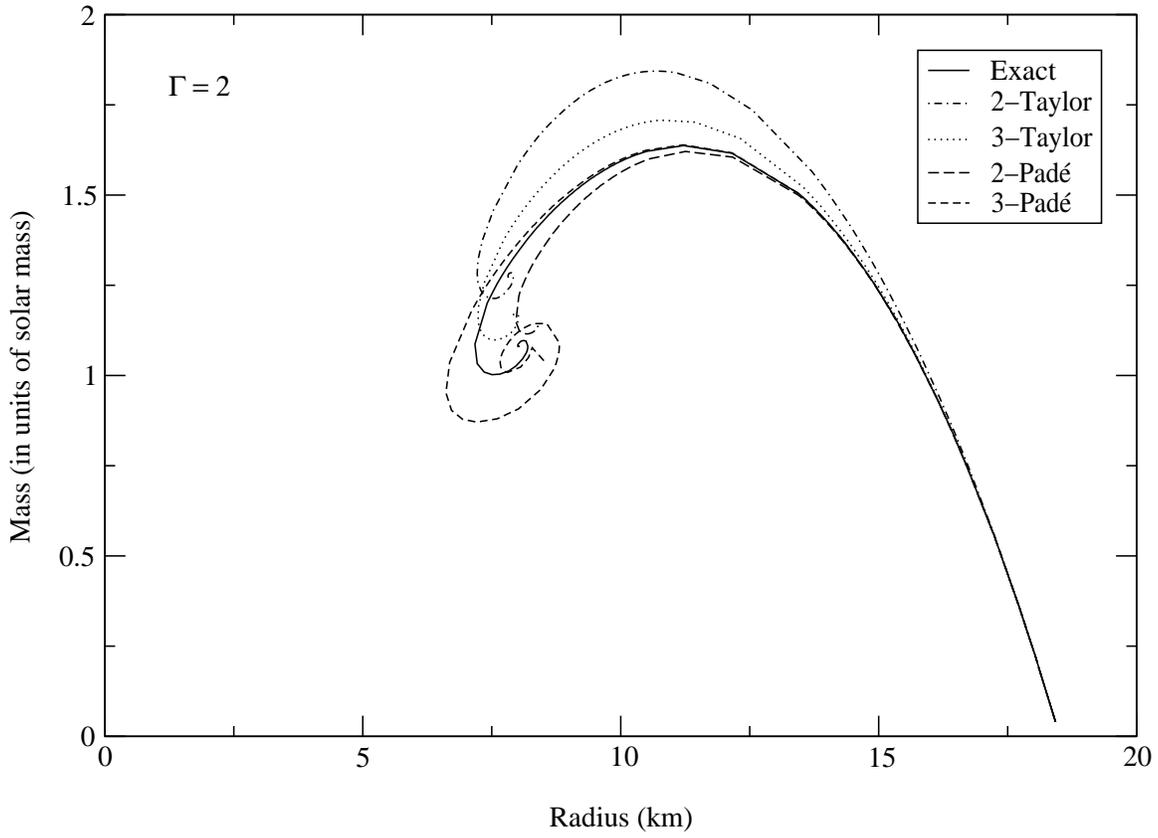}
\bigskip

\caption{Total mass ($M_{\odot}$ units) as a function of
  radius (km) for the intermediate value of $\Gamma$ discussed in
  Fig.~\ref{fig:f1} i.e., $\Gamma = 2$.  The deviation of the 3-Pad\'e
  curve with respect to the GR curve for smaller radii models correspond
  to models (with high values of the central density) that fall in the
  unstable branch of the mass-central density curves. 
  Pad\'e models should not be used beyond the maximum mass limit.}
\label{fig:f2}
\end{figure}

\begin{figure}
\epsfxsize=6in \epsfbox{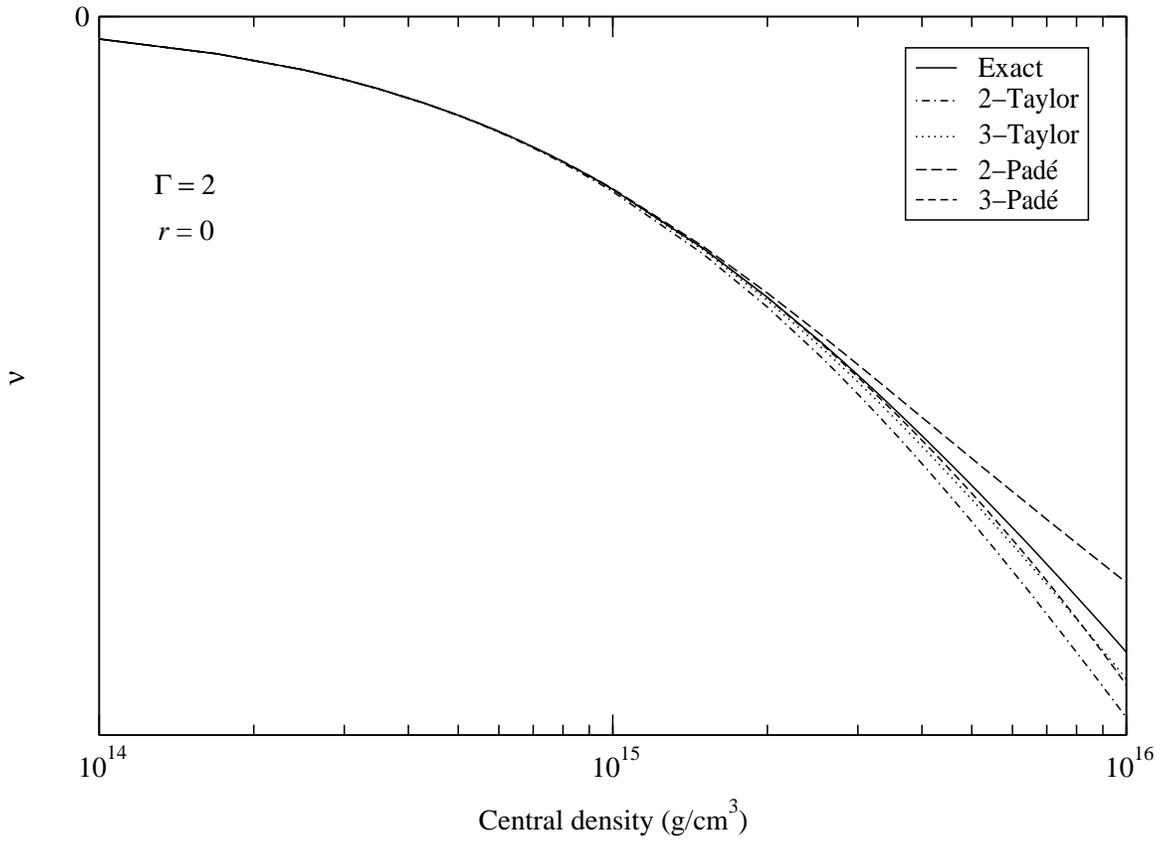}
\bigskip

\caption{The exponent $\nu$ of metric component $g_{t t}$ at 
  $r=0$ as a function of central density.}
\label{fig:f3}
\end{figure}

In Fig.~\ref{fig:f1}, we plot mass (in solar mass units) as
a function of central density. 
The different curves represent the exact TOV,  
the 2PN and 3PN  truncated (Taylor) models
  (henceforth 2-Taylor, 3-Taylor etc; for more details, see \cite{shinkai}) 
and our new Pad\'e approximated models. 
We do not  plot curves corresponding to
 1PN truncated Taylor and the associated  
first order Pad\'e model  as they are very far
away from the exact GR curve and thus evidently  inadequate. 
It is very clear from the figure that for all models to the left
 of the (first) maximum, 
which represents the stable branch of the mass-density curve,
the Pad\'e models do extremely well as compared to Taylor models. 
In this regime the  third order Pad\'e (now onwards 3-Pad\'e) curve (dashed line) 
is very close to
the exact GR curve (solid line). The 2-Pad\'e (long-dashed line) curve is not 
only better compared to the 2-Taylor one (dot-dashed) but even far better than 
the 3-Taylor curve (dotted line). In the unstable branch i.e., to
the right to the maximum, the Pad\'e approximation starts becoming bad.
This feature is more severe when
the stiffness of the considered equation of state and the order of the
Pad\'e approximants are higher\footnote{While computing
the PN coefficients in Eq. (\ref{eq:dpr3}) for the models which fall in the
unstable branch, we notice that the values for the 1PN terms become even 
greater than one, indicating that neither the Taylor truncation nor the 
Pad\'e expansion is reliable in this region. For the stable branch models 
these coefficients are always lesser than one as shown in Fig.~\ref{fig:r1}. }
(Fig.~\ref{fig:f1}b and c).  However,  the stable branch
is approximated well by Pad\'e models even for a very stiff 
equation of state ($\Gamma = 3$, Fig.~\ref{fig:f1}c), 
where the approximation breaks down near ($2.75$ x $10^{15}$ g/cm$^3$)
the maximum mass limit ($1.85$ x $10^{15}$ g/cm$^3$).
Studies of mass-radius curves and $\nu_c$ (the exponent of the metric component
 $g_{t t}$ at the center of the star) {\it vs} central density curves also
show a similar dependence on the stiffness of the chosen equation of state 
and on the 
central density. In Fig.~\ref{fig:f2} and Fig.~\ref{fig:f3}, we plot 
mass {\it vs} radius and $\nu_c$ {\it vs} central density curves respectively,
for an equation of state with intermediate stiffness $\Gamma = 2$.
The deviation of the 3-Pad\'e solutions from the exact GR ones correspond to 
the configurations lying on the unstable branch. 

\begin{figure}
\epsfxsize=6in \epsfbox{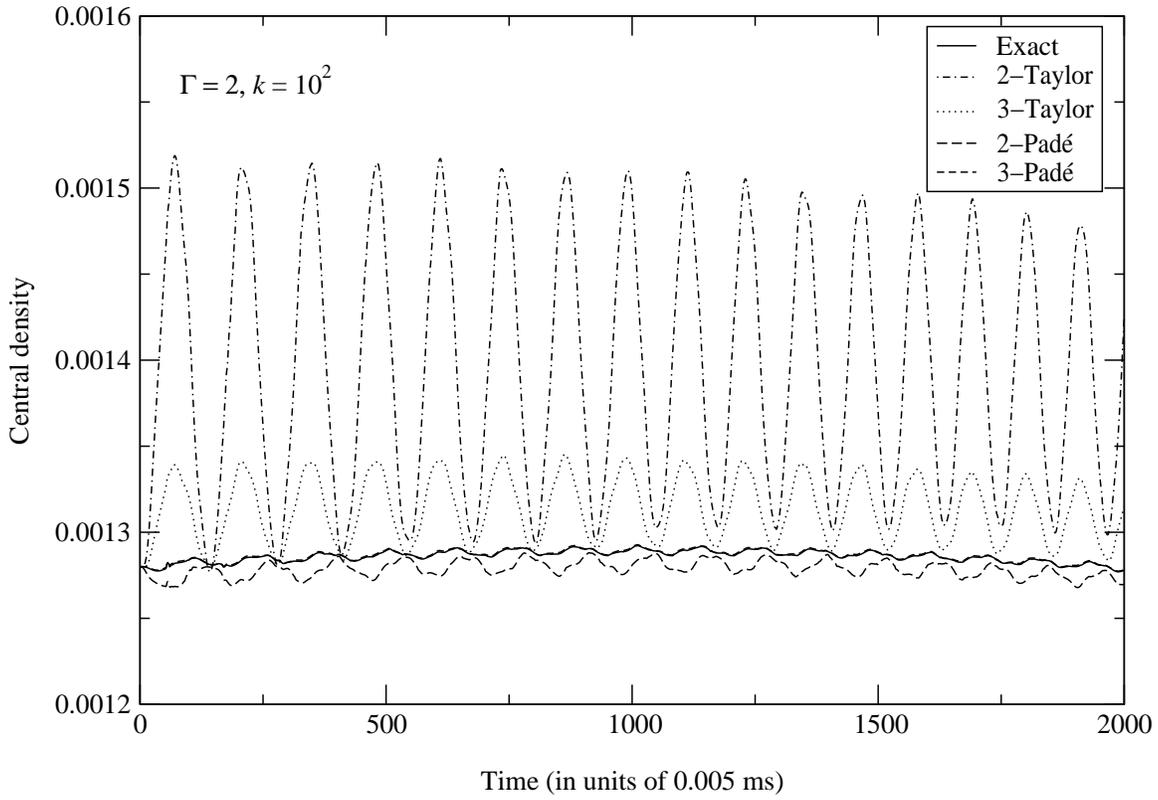}
\bigskip

\caption{The evolution of central density for exact TOV, Pad\'e 
  models and Taylor truncated TOV initial configurations with the same value of
  central density $\rho_c = 1.28$ x $10^{-3}$ and $\Gamma = 2$; $ k =
  10^2$.  The numerical truncation error triggers the dynamical
  evolution and shows the pulsation which corresponds to the physical
  pulsation modes of a stable spherically symmetric model.}
\label{fig:f4}
\end{figure}

\begin{figure}
\epsfxsize=6in \epsfbox{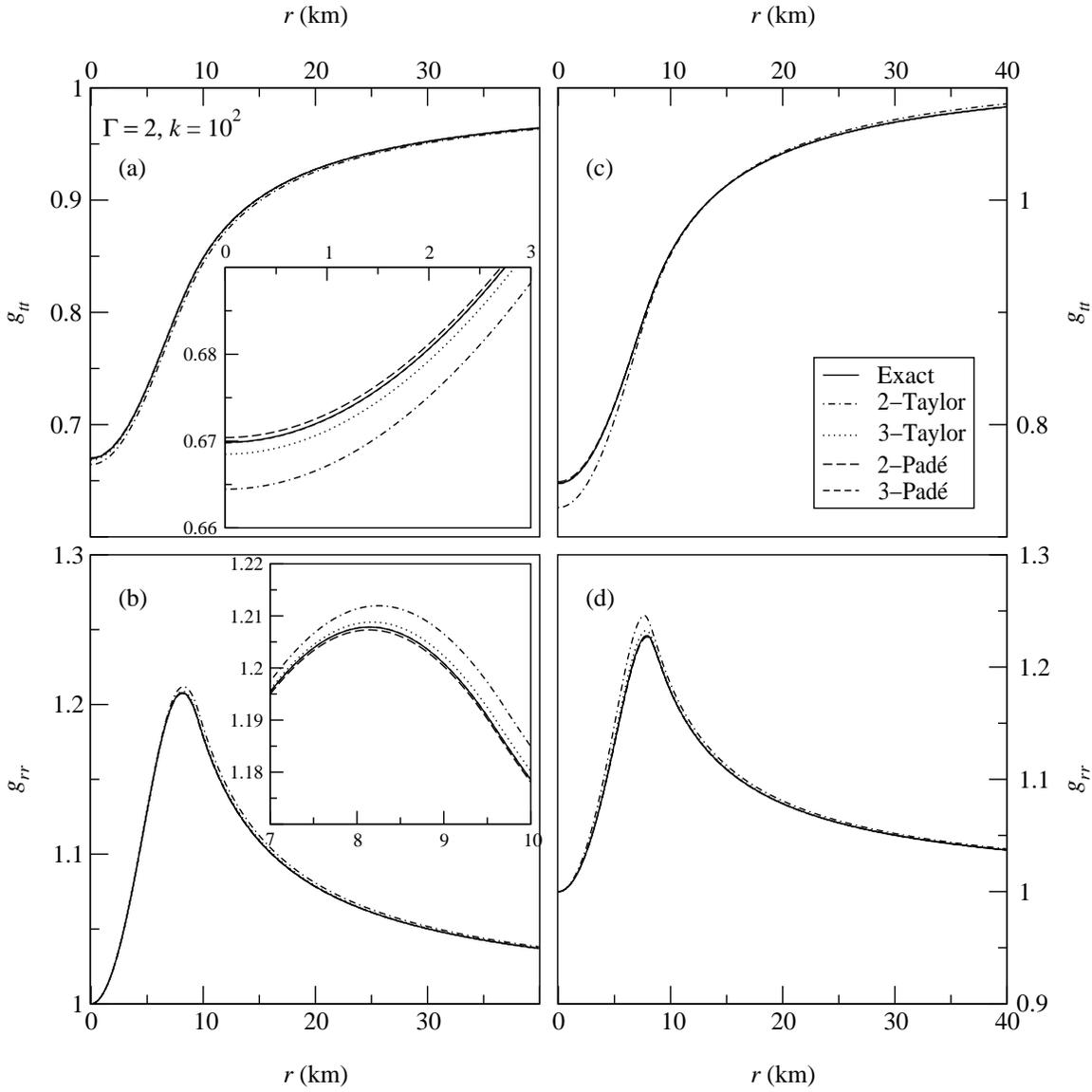}
\bigskip

\caption{$g_{t t}$ and $g_{r r}$ over the grid at time $t = 0$\,ms 
  ((a) and (b), respectively).  The insets zoom into particular
  portions of the grid (near the center for $g_{t t}$ and near the
  surface for $g_{r r}$) to highlight the Pad\'e behaviour more
  clearly. (c) and (d) are the same as (a) and (b), respectively, but
  at $t = 10$\,ms.  The model is the same as in Fig.~\ref{fig:f4}.}
\label{fig:f5}
\end{figure}

\begin{figure}
\epsfxsize=6in \epsfbox{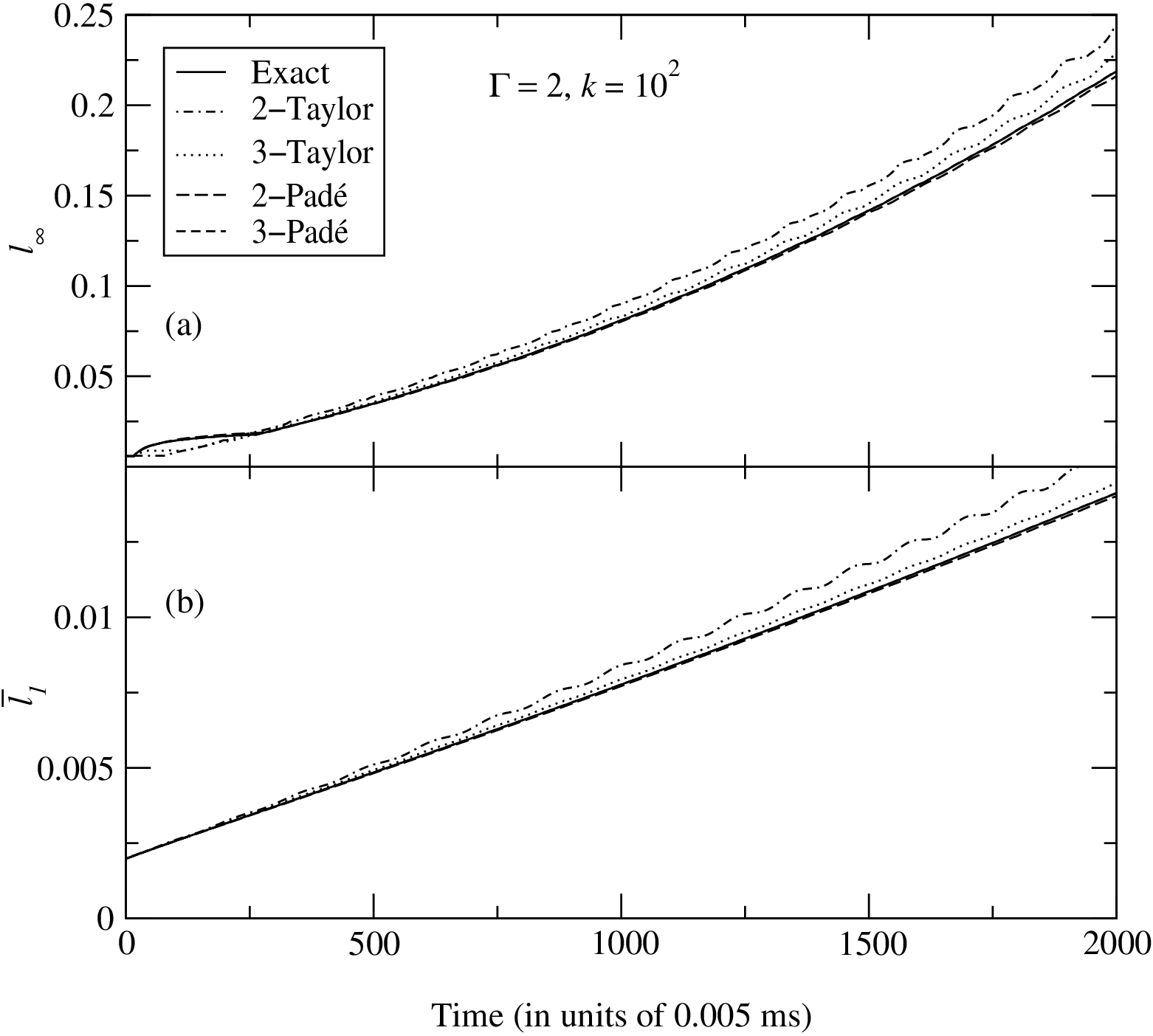}
\bigskip

\caption{The norms of the Hamiltonian constraint equation 
  $l_{\infty}$ (maximum value at a given time step) and
  ${\bar l}_1$ (the average of the absolute value) as a function
  of time ((a) and (b), respectively). The model is the same as in
  Fig.~\ref{fig:f4}.  }
\label{fig:f6}
\end{figure}

The studies of these equilibrium configurations imply clearly that the 
stable general relativistic TOV configurations are approximated very well by
Pad\'e models as compared to Taylor truncated models, even if
they do not perform so well in the unstable branch. To study 
this  further, we next  numerically
 evolve these five initial configurations - Exact TOV, 2-Pad\'e, 3-Pad\'e, 
2-Taylor and 3-Taylor truncated models - in time and qualitatively compare 
their evolution. 
We use a spherically
symmetric general relativistic hydrodynamical code for this purpose 
\cite{credit}.
The code uses polar slicing and radial gauge. The spacetime is 
evolved using the ADM formalism and the hydrodynamic evolution
is based on a high resolution shock 
capturing \cite{toni}-\cite{miller} scheme. The grid boundary is fixed at 
about four times the radius of the TOV model.  We choose a model with 
central density $1.28$ x $10^{-3}$ (geometrised units), $\Gamma = 2 , 
k = 10^2$ and then evolve it upto 10 ms (though the code runs for a much 
longer time, 
10 ms evolution is sufficient for our analysis).
In Fig.~\ref{fig:f4} 
the evolution of central density using various initial data is displayed. 
 We find  that the 3-Pad\'e curve follows the TOV
evolution curve very closely.
 Even the 2-Pad\'e evolution is  much better than the 3-Taylor 
and the 2-Taylor evolutions.
The oscillations in the central density are due to the numerical truncation
errors which introduce non-zero radial velocity. These truncation errors
act as the small perturbations on a stable spherically symmetric configuration 
and give rise to the radial pulsation modes of the system. 
 A Fourier transform
of the density or radial velocity evolution can be used to extract these
pulsation modes \cite{nick} and we hope to return to this in a subsequent work. 
In Fig.~\ref{fig:f5} we plot the metric components $g_{t t}$ and $g_{r r}$ 
over the grid for $t = 0$ (initial time) and $t = 10$\,ms
(final time). 
To compare the global performance of the initial data used (the exact
TOV and the various truncated models),
the norms of the Hamiltonian constraints $l_{\infty}$ (maximum value
at a given time step) and ${\bar l}_1$ (average of the absolute value) 
are evaluated and shown in Fig.~\ref{fig:f6}.
These figures, once again, confirm the superior performance of Pad\'e 
approximants over 
Taylor approximants of the same order. Though we do not display them,
 we get similar results for the other stable configurations
both for  relatively softer and  stiffer equations of state.
  A final comment on the models in the unstable branch
 of the mass-density curve:
the truncation errors are enough to trigger a collapse and make the system
unstable on a   time scale  so short that a comparison of various
approximations is not possible. A more detailed analysis would be
needed in this regard.

To conclude: Detailed studies of
equilibrium configurations of single neutron stars and their evolution 
indicate, that in the stable branch, the  second order Pad\'e
 model converges to the exact general relativistic model
even  better than the  straightforward third order truncated
 (Taylor) PN  model. 
 Both the  simpler one parameter  Pad\'e form
and a more involved three parameter Pad\'e form  exhibit 
similar improvement over the Taylor models. 
The Pad\'e models are thus quite robust and controlled  
and  perform better than the simpler Taylor truncated models.
It is better to use  initial data obtained from
 a Pad\'e approximant to the  Taylor model
 than initial data  from a straightforward post-Newtonian truncated 
model of the same order. This feature should be generic and extend to 
binary neutron stars and black holes [especially since a useful simplification
in a two-body problem is via a reduction to an equivalent one-body
problem] and prove useful in numerical studies
of such systems in the future. 

\acknowledgements

A. Gopakumar is supported in part by the National Science Foundation
Grants No. PHY 96-00049, 96-00507 and 99-79985 and NASA Grant No. NCCS 5-153.  
We thank the AEI-WashU numerical relativity
collaboration, in particular J. A. Font, T. Goodale, M. Miller, E. Seidel, 
N.Stergioulas and  W-M. Suen, for providing the evolution code
and for useful clarifications and discussions. We also thank H.
Shinkai for the helpful discussion regarding the Hamiltonian
constraint solver. B. R. Iyer thanks M. Campanelli and C. Lousto for
discussions. A. Gupta thanks TPSC (Theoretical Physics Seminar
Circuit) for supporting a visit to PRL, Ahmedabad, during which this
work was started.
Finally, we thank the referee for critical and
valuable comments that helped clarify the construction and  presentation.

\appendix

\section{}

The 3PN truncated model in general has the form
\begin{eqnarray}
T_3 & = & 1 + t_A A + t_B B + t_C C + \nonumber\\
    &&t_{A A} A^2 + t_{B B} B^2 + t_{C C} C^2 + t_{A B} A B + t_{B C} B C
+ t_{C A} C A + \nonumber\\
&&  t_{AA A} A^3 + t_{BB B} B^3 + t_{CC C} C^3+  t_{ABC}ABC+\nonumber\\
    &&t_{A AB} A^2B + t_{B BC} B^2C + t_{C CA} C^2A + t_{A BB} A B^2 +
t_{B CC} B C^2 +t_{A AC} A^2 C \,.
\label{eq:tayl3}
\end{eqnarray}
The associated 3PN Pad\'e may be written as       

\begin{mathletters}
\begin{eqnarray}
P_3&=&\frac{1}{F_3},\\
{\rm where }\; F_3&=& 1 +
\frac{p_1 A}{1+\frac{p_{1 1} A}{S_{11}}
 + \frac{p_{1 2} B}{S_{12}} + \frac{p_{1 3} C}{S_{13}} }+\nonumber \\
&&\frac{p_2 B}{1+\frac{p_{2 1} A}{S_{21}}
 + \frac{p_{2 2} B}{S_{22}} + \frac{p_{2 3} C}{S_{23}}} +
\frac{p_3 C}{1+\frac{p_{3 1} A}{S_{31}}
 + \frac{p_{3 2} B}{S_{32}} + \frac{p_{3 3} C}{S_{33}}} \,,
\end{eqnarray}
\end{mathletters}
where $S_{ij} = 1+p_{ij1}A+p_{ij2}B+p_{ij3}C$, $i,j=1\ldots 3$.
The analysis discussed in Sec. III at 2PN may be repeated
at 3PN after making the following natural  symmetric choice         
implied by the requirement that all Pad\'e coefficients contributing
to a particular Taylor term contribute equally:
\begin{mathletters}
\label{3pnsym}
\begin{eqnarray}
p_{112}&=& p_{121}=p_{211} \,,\\
p_{122}&=&p_{221}=p_{212}\,,\\
p_{113}&=&p_{131}=p_{311}\,,\\
p_{123}&=&p_{132}=p_{231}=p_{213}=p_{312}=p_{321}\,,\\
p_{313}&=&p_{331}=p_{133}\,,\\
p_{322}&=&p_{232}=p_{223}\,,\\
p_{332}&=&p_{233}=p_{323}\,.
\end{eqnarray}
\end{mathletters}
The remaining $10$ independent $3$PN Pad\'e coefficients, uniquely
determined by 1PN, 2PN and 3PN Taylor coefficients, are given by
\begin{mathletters}
\label{3pnsol}
\begin{eqnarray}
{\it p_{111}}&=&{\frac {-{t_{{{\it AA}}}}^{2}+t_{{{\it AAA}}}t_{{A}}}{t_{{
A}}\left ({t_{{A}}}^{2}-t_{{{\it AA}}}\right )}}\,,
%%%%%%
\\
{\it p_{222}}&=&{\frac {-{
t_{{{\it BB}}}}^{2}+t_{{{\it BBB}}}t_{{B}}}{t_{{B}}\left ({t_{{B}}}^{2
}-t_{{{\it BB}}}\right )}}\,,
%%%%%
\\
{\it p_{333}}&=&{\frac {-{t_{{{\it CC}}}}^{2}+t_
{{{\it CCC}}}t_{{C}}}{t_{{C}}\left ({t_{{C}}}^{2}-t_{{{\it CC}}}
\right )}}\,,
\\
{\it p_{112}}&=& \le \biggl [-{t_{{A}}}^{4}t_{{B}}
+2\,{t_{{B}}}^{2}{t_{{A}}}^{3}
+\left (-{t_{{B}}}^{3}+\left (-2\,t_{{{\it AB}}}+2\,t_{{{\it AA}}}
\right )t_{{B}}+t_{{{\it AAB}}}\right ){t_{{A}}}^{2} 
\no 
&&
+\left (-2\,t_{{{
\it AA}}}t_{{{\it AB}}}+2\,t_{{{\it AAB}}}t_{{B}}+2\,{t_{{B}}}^{2}t_{{
{\it AB}}}\right )t_{{A}}-2\,t_{{{\it AA}}}{t_{{B}}}^{3}+t_{{{\it AAB}
}}{t_{{B}}}^{2}
\no
&&
-t_{{{\it AB}}}\left (2\,t_{{{\it AA}}}+t_{{{\it AB}}}
\right )t_{{B}}\biggr ] \ri 
\no
&&
\le \left (t_{{B}}+t_{{A}}\right )^{2}\left (2\,t_{
{A}}t_{{B}}-t_{{{\it AA}}}-t_{{{\it AB}}}+{t_{{A}}}^{2}\right )\ri ^{-1} \,,
\\
%%%%%%%%%%%%%
{\it p_{122}}&=& \le \biggl [\left (-{t_{{B}}}^{2}-2\,t_{{{\it BB}}}\right )
{t_{{A}}}^{3}+
\left (2\,t_{{B}}t_{{{\it AB}}}+2\,{t_{{B}}}^{3}+t_{{{\it ABB}}}
\right ){t_{{A}}}^{2}
\no
&&
+\left (-{t_{{B}}}^{4}+\left (2\,t_{{{\it BB}}}-2
\,t_{{{\it AB}}}\right ){t_{{B}}}^{2}+2\,t_{{{\it ABB}}}t_{{B}}-t_{{{
\it AB}}}\left (t_{{{\it AB}}}+2\,t_{{{\it BB}}}\right )\right )t_{{A}}
\no
&&
-2\,t_{{{\it AB}}}t_{{{\it BB}}}t_{{B}}+t_{{{\it ABB}}}{t_{{B}}}^{2}
\biggr ] \ri
\no
&&
\le \left (t_{{B}}+t_{{A}}\right )^{2}\left ({t_{{B}}}^{2}+2\,t_{{
A}}t_{{B}}-t_{{{\it AB}}}-t_{{{\it BB}}}\right ) \ri ^{-1}\,,
\\
%%%%%%%%%%%%%%%
{\it p_{113}}&=& \le \biggl [-t_
{{C}}{t_{{A}}}^{4}+2\,{t_{{A}}}^{3}{t_{{C}}}^{2}+\left (-{t_{{C}}}^{3}
+\left (2\,t_{{{\it AA}}}-2\,t_{{{\it AC}}}\right )t_{{C}}+t_{{{\it 
AAC}}}\right ){t_{{A}}}^{2}
\no
&&
+\left (-2\,t_{{{\it AA}}}t_{{{\it AC}}}+2
\,{t_{{C}}}^{2}t_{{{\it AC}}}+2\,t_{{{\it AAC}}}t_{{C}}\right )t_{{A}}
-2\,t_{{{\it AA}}}{t_{{C}}}^{3}+t_{{{\it AAC}}}{t_{{C}}}^{2}
\no
&&
-t_{{{\it AC}}}
\left (t_{{{\it AC}}}+2\,t_{{{\it AA}}}\right )t_{{C}}\biggr ]\ri
\no
&&
\le \left (t_{{A}}+t_{{C}}\right )^{2}\left ({t_{{A}}}^{2}+2\,t_{{C}}t_{{A
}}-t_{{{\it AA}}}-t_{{{\it AC}}}\right ) \ri^{-1}\,,
\\
%%%%%%%%%%%%%%%%%%%%%
{\it p_{123}}&=&\le \biggl [\left (-2\,{t_{{B}}}^{2}t_{{C}}
+\left (-2\,{t_{{C}}}^{2}
-2\,t_{{{\it BC}}}\right )t_{{B}}-2\,t_{{{\it BC}}}t_{{C}}\right 
){t_{{A}}}^{3}
\no
&&
+\biggl ( -2\,t_{{C}}{t_{{B}}}^{3}+\left (2\,t_{{{\it BC}}
}+2\,t_{{{\it AC}}}+12\,{t_{{C}}}^{2}\right ){t_{{B}}}^{2}
\no
&&
+\left (-2\,
{t_{{C}}}^{3}+\left (2\,t_{{{\it AB}}}-4\,t_{{{\it BC}}}+2\,t_{{{\it 
AC}}}\right )t_{{C}}+t_{{{\it ABC}}}\right )t_{{B}}
\no
&&
+\left (2\,t_{{{
\it AB}}}+2\,t_{{{\it BC}}}\right ){t_{{C}}}^{2}+t_{{{\it ABC}}}t_{{C}
}\biggr ){t_{{A}}}^{2}
\no
&&
+\biggl (\left (-2\,{t_{{C}}}^{2}-2\,t_{{{\it AC}
}}\right ){t_{{B}}}^{3}+\left (-2\,{t_{{C}}}^{3}+\left (2\,t_{{{\it AB
}}}+2\,t_{{{\it BC}}}-4\,t_{{{\it AC}}}\right )t_{{C}}+t_{{{\it ABC}}}
\right ){t_{{B}}}^{2}
\no
&&
+\left (\left (2\,t_{{{\it AC}}}+2\,t_{{{\it BC}}
}-4\,t_{{{\it AB}}}\right ){t_{{C}}}^{2}+2\,t_{{{\it ABC}}}t_{{C}}-2\,
t_{{{\it AB}}}\left (t_{{{\it BC}}}
+t_{{{\it AC}}}\right )\right )t_{{ B}}
\no
&&
-2\,{t_{{C}}}^{3}t_{{{\it AB}}}+t_{{{\it ABC}}}{t_{{C}}}^{2}-2\,t_{
{{\it AC}}}\left (t_{{{\it AB}}}
+t_{{{\it BC}}}\right )t_{{C}}\biggr )
t_{{A}}
\no
&&
-2\,t_{{C}}{t_{{B}}}^{3}t_{{{\it AC}}}+\left (\left (2\,t_{{{
\it AC}}}+2\,t_{{{\it AB}}}\right ){t_{{C}}}^{2}+t_{{{\it ABC}}}t_{{C}
}\right ){t_{{B}}}^{2}
\no
&&
+\left (-2\,{t_{{C}}}^{3}t_{{{\it AB}}}+t_{{{
\it ABC}}}{t_{{C}}}^{2}-2\,t_{{{\it BC}}}\left (t_{{{\it AB}}}+t_{{{
\it AC}}}\right )t_{{C}}\right )t_{{B}}\biggr ] \ri
\no
&&
\le \left (t_{{A}}+t_{{C}}
\right )\left (t_{{B}}+t_{{C}}\right )
\left (t_{{B}}+t_{{A}}\right )
\left (2\,t_{{A}}t_{{B}}+2\,t_{{B}}t_{{C}}+2\,t_{{C}}t_{{A}}-t_{{{\it 
AC}}}-t_{{{\it AB}}}-t_{{{\it BC}}}\right ) \ri^{-1}\,,
\\
%%%%%%
{\it p_{133}}&=&\le \biggl [\left (-
2\,t_{{{\it CC}}}-{t_{{C}}}^{2}\right ){t_{{A}}}^{3}+\left (2\,{t_{{C}
}}^{3}+2\,t_{{C}}t_{{{\it AC}}}+t_{{{\it ACC}}}\right ){t_{{A}}}^{2}
\no
&&
+\left (-{t_{{C}}}^{4}+\left (2\,t_{{{\it CC}}}-2\,t_{{{\it AC}}}
\right ){t_{{C}}}^{2}+2\,t_{{{\it ACC}}}t_{{C}}-t_{{{\it AC}}}\left (2
\,t_{{{\it CC}}}+t_{{{\it AC}}}\right )\right )t_{{A}}
\no
&&
+t_{{{\it ACC}}}
{t_{{C}}}^{2}-2\,t_{{{\it AC}}}t_{{{\it CC}}}t_{{C}}\biggr ] \ri
\no
&&
\le \left (t_{A}
+t_{{C}}\right )^{2}\left (2\,t_{{C}}t_{{A}}-t_{{{\it CC}}}-t_{{{
\it AC}}}+{t_{{C}}}^{2}\right ) \ri^{-1} \,,
\\
%%%%%%%%%%
{\it p_{223}}&=& \le \biggl [-t_{{C}}{t_{{B}}}^{4
}+2\,{t_{{B}}}^{3}{t_{{C}}}^{2}+\left (-{t_{{C}}}^{3}+\left (2\,t_{{{
\it BB}}}-2\,t_{{{\it BC}}}\right )t_{{C}}
+t_{{{\it BBC}}}\right ){t_{ {B}}}^{2}
\no
&&
+\left (-2\,t_{{{\it BB}}}t_{{{\it BC}}}+2\,{t_{{C}}}^{2}t_{{
{\it BC}}}+2\,t_{{{\it BBC}}}t_{{C}}\right )t_{{B}}
\no
&&
-2\,t_{{{\it BB}}}{
t_{{C}}}^{3}+t_{{{\it BBC}}}{t_{{C}}}^{2}-t_{{{\it BC}}}\left (2\,t_{{
{\it BB}}}+t_{{{\it BC}}}\right )t_{{C}}\biggr ] \ri
\no
&&
\le \left (t_{{B}}+t_{{C}}
\right )^{2}\left ({t_{{B}}}^{2}+2\,t_{{B}}t_{{C}}-t_{{{\it BC}}}-t_{{
{\it BB}}}\right ) \ri^{-1}\,,
\\
%%%%%%%%%%%%%
{\it p_{233}}&=& \le \biggl [\left (-2\,t_{{{\it CC}}}-{t_{{C}
}}^{2}\right ){t_{{B}}}^{3}+\left (t_{{{\it BCC}}}+2\,{t_{{C}}}^{3}+2
\,t_{{{\it BC}}}t_{{C}}\right ){t_{{B}}}^{2}
\no
&&
+\left (-{t_{{C}}}^{4}+
\left (2\,t_{{{\it CC}}}-2\,t_{{{\it BC}}}\right ){t_{{C}}}^{2}+2\,t_{
{{\it BCC}}}t_{{C}}-t_{{{\it BC}}}\left (2\,t_{{{\it CC}}}+t_{{{\it BC
}}}\right )\right )t_{{B}}
\no
&&
+t_{{{\it BCC}}}{t_{{C}}}^{2}-2\,t_{{{\it BC
}}}t_{{{\it CC}}}t_{{C}}\biggr ] \ri
\no
&&
\le \left (t_{{B}}+t_{{C}}\right )^{2}
\left (2\,t_{{B}}t_{{C}}-t_{{{\it CC}}}+{t_{{C}}}^{2}-t_{{{\it BC}}}
\right ) \ri^{-1} \,.
\end{eqnarray}
\end{mathletters}
[The above solution obtains if none of the factors in the denominators of the
above expressions are vanishing.]

For the $F_3$ associated with the TOV equation at 3PN order, we have
\begin{mathletters}
\begin{eqnarray}
t_{CCC}&=&t_{ABC}=t_{ACC}=t_{BCC}=1\,,\\
t_{AAA}&=&t_{BBB}=t_{AAB}=t_{ABB}=t_{BBC}=t_{AAC}=0\,.
\end{eqnarray}    
\end{mathletters}

\begin{mathletters}
\label{3pnsymtov}
\begin{eqnarray}
p_{111}&=& p_{222}=0\,,\\
p_{112}&=& p_{121}=p_{211}=-{1 \over 8} \,,\\
p_{122}&=&p_{221}=p_{212}=-{ 1 \over 8}\,,\\
p_{113}&=&p_{131}=p_{311}=-{1 \over 8}\,,\\
p_{123}&=&p_{132}=p_{231}=p_{213}=p_{312}=p_{321} =-{1 \over 6}\,,\\
p_{313}&=&p_{331}=p_{133} =-{1 \over 4}\,,\\
p_{322}&=&p_{232}=p_{223}=-{1 \over 8}\,,\\
p_{332}&=&p_{233}=p_{323}=-{1 \over 4}\,.
\end{eqnarray}
\end{mathletters}
The solution  for $p_{333}$, Eq. (A4c), is not applicable, since 
$t_C=t_{CC}=1$. One finds that $p_{333}$ is
indeterminate. We  choose  it to be zero.
The function $F_3$ and $G_3$ then  reduce to     

\begin{mathletters}
\begin{eqnarray}
F_3&=&1- \frac{A}{D_1} - \frac{B}{D_2} - \frac{C}{D_3}\,,\\
%NF33
D_1 &=&1+{A \over (1-{(B+C)\over 8})}
+{ B \over 2\,(1-{(A +B) \over 8}-{C \over 6} ) }
+{ C \over 2\,(1-{ A \over 8}-{ B \over 6}-{C \over 4})}\,,\\
%NF32
D_2 &=&1+{ A \over 2 \,( 1-{(A +B) \over 8}-{C \over 6})}
+{B \over 1-{(A+C) \over 8}}
+{C \over 2\,(1-{A\over 6}-{B \over 8}-{C \over 4} )}\,,\\
%NF31
D_3 &=&1+{ A \over 2\,(1-{A \over 8}-{ B \over 6}-{C \over 4})}
+{ B \over 2\,( 1-{ A \over 6}-{B \over 8 }-{C \over 4})}\,.
\end{eqnarray} 
\end{mathletters}

\begin{mathletters}
\begin{eqnarray}
G_3&=& 1 - \frac{B}{D_5} - \frac{C}{D_6}\,,\\
D_5&\equiv&1+
\frac{B}{1-\frac{C}{8}}+
\frac{C}{2\,(1-\frac{B}{8}-\frac{C}{4})},\\
D_6&\equiv&1+
\frac{B}{2\,(1-\frac{B}{8}-\frac{C}{4})}\,.
\end{eqnarray}   
\end{mathletters}

The corresponding  one parameter Pad\'e approximant at 3PN order
on the other hand is given by
\begin{mathletters}
\begin{eqnarray}
{\cal F}_3&=&
\left\{A^3+(1+B+C)A^2+[B^2+(1+C)B+C]A+B^3+(1+C)B^2+BC\right\}^{-1}\nonumber\\
&&\le [B^2+(C-1)B-C+1]A^2+[(C-1)B^2+(-2C+C^2+1)B-C^2+C]A+\nonumber\\
&&\;\;\;(1-C)B^2+(C-C^2)B\ri \,,\\
{\cal G}_3&=&\frac
{(1-C)B^2+(C-C^2)B}
{B^3+(1+C)B^2+BC}\,.
\end{eqnarray}
\end{mathletters}

\newpage

\newpage

\newpage
\newpage
\newpage
\newpage
\newpage
\newpage
\newpage


\begin{references}
\bibitem{credit}
The code was originally developed by Mark Miller, Washington University.
A version modified by Tom Goodale, AEI, was used in this work.
\end{references}
\end{document}